\documentclass[%
aps,
prl,
reprint,
amsmath,
amssymb,
amsfonts,
superscriptaddress,
]{revtex4-2}
\usepackage{graphicx}
\usepackage{xcolor}
\usepackage{bm}
\usepackage[%
hidelinks,
colorlinks=true,
citecolor=blue,
urlcolor=blue,
linkcolor=blue
]{hyperref}
\usepackage[inline]{enumitem}

\definecolor{red}{HTML}{C74431}

\definecolor{green}{HTML}{3C9455}

\begin{document}
\author{Roland Wiese}
\email{wiese@itp.uni-leipzig.de}
\author{Klaus Kroy}
\affiliation{Institute for Theoretical Physics, Leipzig University, 04103 Leipzig, Germany}
\author{Demian Levis}
\email{levis@ub.edu}
\affiliation{Departement de F{\'i}sica de la Materia Condensada, Facultat de F{\'i}sica, Universitat de Barcelona, Mart{\'i} i Franqu{\`e}s 1, 08028 Barcelona, Spain}
\affiliation{University of Barcelona Institute of Complex Systems (UBICS), Facultat de F{\'i}sica,  Universitat de Barcelona, Mart{\'i} i Franqu{\`e}s 1, 08028 Barcelona, Spain}
\date{\today}
\title{Fluid-Glass-Jamming Rheology of Soft Active Brownian Particles}
\begin{abstract}
  We numerically study the shear rheology of a binary mixture of soft Active Brownian Particles, from the fluid to the disordered solid regime.
  At low shear rates, we find a Newtonian regime, where a Green-Kubo relation with an effective temperature provides the linear viscosity.
  It is followed by a shear-thinning regime at high shear rates.
  At high densities, solidification is signalled by the emergence of a finite yield stress.
  We construct a ``fluid-glass-jamming'' phase diagram with activity replacing temperature. 
  While both parameters gauge
 fluctuations, 
 activity also changes the exponent characterizing the decay of the diffusivity close to the glass transition and the shape of the yield stress surface. The dense disordered active solid  appears to be mostly dominated by athermal jamming rather than glass rheology.
\end{abstract}

\maketitle

Ensembles of repulsive particles 
commonly undergo a phase transition  from a fluid to a solid state upon compression.
If the tendency to crystallize is frustrated (e.g. by size polydispersity), thermal systems exhibit a glass transition to a disordered solid \cite{pusey1986phase-behaviour-of-concentrated-suspensions, mewis2011colloidal-suspension-rheology}. 
Solidity can also emerge upon compression in athermal particle systems, such as foams or grains: the so-called jamming transition \cite{nagel2010jamming-transition-and-marginally-jammed-solid}.
Both transitions share the existence of a critical density beyond which a yield stress emerges, heralded by a dramatic slowing down of the dynamics \cite{berthier2011theoretical, bonn2017yield-stress-materials}.
Accordingly, a unified picture in terms of a ``jamming'' phase diagram was proposed \cite{liu1998jamming-is-not-just-cool-any-more, trappe2001jamming-phase-diagram-for-attractive-particles}.
The yield stress surface  has since  been quantified with the help of idealized  particle models and rheological experiments \cite{ciamarra2010recent-results-on-the-jamming-phase-diagram, liu2011universal-jamming-phase-diagram-in-the-hard-sphere-limit, ikeda2012unified, ikeda2013disentangling-glass-and-jamming}, helping to decipher and classify the mechanisms responsible for the emergence of rigidity in diverse soft materials \cite{bonn2017yield-stress-materials}. 

\begin{figure}[t]
  \centering
  \includegraphics[width=\linewidth]{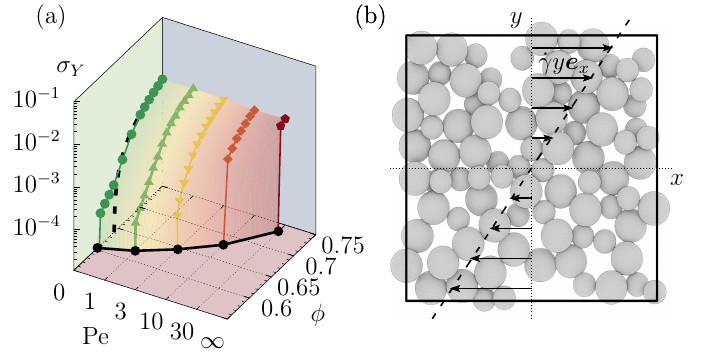}
  \caption{Active ``glass-jamming" rheology: (a) Yield stress surface of harmonic ABPs at reduced temperature $T=10^{-4}$.
    The P{\'e}clet number $\mathrm{Pe}$ is a dimensionless measure of the active against the passive (thermal) particle motility and $\phi$ is the volume fraction.
    The dashed black line corresponds to the athermal ($T=0$) passive jamming limit.
    The black symbols in the $\mathrm{Pe}-\phi$ plane locate the active glass transition at $\phi_G(\mathrm{Pe})$.
    (b) Cross section in the $xy$-plane of the 3-dimensional simulation box, showing the imposed shear velocity profile. 
  } \label{fig:phasediagram}
\end{figure}

A resurgence of interest in understanding the emergence of solidity is recently observed in an \emph{a priori} completely new context, namely, dense disordered active matter.
Indeed, assemblies of biological cells \cite{angelini2011glass-like-dynamics-of-collective-cell-migration, manning2013glassy-dynamics-in-3d-embryonic-tissues} and synthetic active colloids \cite{klongvessa2019active-glass-ergodicity-breaking} display a fluid to solid transition, key to understanding tissue mechanics and morphogenesis.
Moreover, they exhibit slow, collective dynamics, reminiscent of supercooled liquids approaching a glass transition. 
Again, this phenomenology has been rationalized in terms of a jamming phase diagram  \cite{mongera2018fluid-to-solid-jamming-transition,manning2021jamming-and-arrest-of-cell-motion-in-biological-tissues, lenne2022sculpting-tissues-by-phase-transitions, campas2022nuclear-jamming-transition-in-vertebrate-organogenesis}, where temperature is replaced by ``activity'', usually in the form of motility, as in Fig.~\ref{fig:phasediagram}(a), where the P{\'e}clet number $\mathrm{Pe}$ quantifies the propulsion velocity of Active Brownian Particles (ABPs). 

While thermal fluctuations can usually be neglected in active systems, the effects of the (nonthermal) active fluctuations are often subsumed into an effective temperature.
It is however conceptually unclear how reliable the qualitative analogy between activity and temperature really is and whether the emergence of a disordered solid should be attributed to a jamming or glass transition. 
The question of how self-propulsion affects a glass transition has already been addressed in numerous works using model systems \cite{berthier2013non-eq-glass-transitions-in-driven-and-active-matter, ni2013pushing-the-glass-transition, berthier2014nonequilibrium-glassy, levis2015single-particle-to-collective-effective-temperatures, berthier2017how-active-forces-influence-noneq-glass-transitions, manning2016motility-driven-glass-and-jamming-transitions-in-biological-tissues, sussman2018anomalous-glassy-dynamics-dense-biological-tissue,  janssen2019active-glasses, manning2019glassy-dynamics-in-models-of-confluent-tissue,  henkes2020dense,sadhukhan2021glassy-dynamics-in-cellular-potts-model, paoluzzi2022mips-to-glassiness-in-dense-active-matter, berthier2022disordered-collective-motion-in-dense-assemblies-of-persistent-particles}.
Active fluctuations have been found to play a crucial role in the unjamming of epithelial tissues \cite{mitchel2020unjamming-transition-distinct-from-EMT} and in gravisensing of plant cells via granular rheology \cite{forterre2018gravisensor-in-plant-cells}.
Yet, despite considerable recent progress both for models of individual ABPs \cite{loewen2011self-propelled-shear-flow} and interacting active many-body systems  \cite{asheichyk2019response, reichert2021transport-coefficients-MCT, mandal2021shear-induced-orientational-ordering,bayram2023motility}, the rheology of dense active matter \cite{peshkov2016active,henkes2017cell-division-and-death-inhibit-glassy-behaviour, amiri2022yielding-transition-in-epithelial-tissues} remains poorly understood, and the distinction between glassy and jamming rheological regimes has not been investigated in this context, so far.
An analogy has been drawn between shear (a global drive) and activity (a local drive) \cite{mo2020rheological-similarities-between-dense-self-propelled-and-sheared-particulate-systems}, and indeed, in infinite spatial dimensions in the infinite-persistence limit, their formal equivalence has been established \cite{morse2021link-active-matter-sheared-granular, agoritsas2021mean-field-dynamics-of-infinite-dimensional-particle-systems}. In two dimensions, the mechanisms that govern yielding in the respective systems were found to be different, though \cite{villarroel2021critical-yielding-rheology}. 
Active disks under shear were reported to orientationally order in the presence of hard walls \cite{mandal2021shear-induced-orientational-ordering} and trigger shear thickening as a result of clustering 
 \cite{bayram2023motility}. 
However, the rheology of dense disordered 3-dimensional ABPs with a finite persistence has not yet been explored.



In this Letter we investigate the shear rheology of self-propelled soft particles by means of computer simulations.
In previous experiments with (dilute) microswimmer suspensions, both hydrodynamic and particle-wall interactions are likely to be crucial \cite{rafai2010effective, gachelin2013non, lopez2015turning, saintillan2018rheology, martinez2020combined-rheometry-and-imaging-study-of-viscosity-reduction-in-bacterial-suspensions}.
Here, in order to decipher the role played by self-propulsion alone,  we consider a simplified model in which none of these two ingredients are at play \cite{shaebani2020computational}.
More precisely, we consider a ``dry'' microswimmer model of harmonic ABPs in three dimensions \cite{wysocki2014cooperative-motion, turci2021phase-separation-and-multibody-effects-in-3D-ABP} with periodic boundary conditions (see Fig.~\ref{fig:phasediagram}(b)).
The restriction to dry active dynamics is quite common in the field, as hydrodynamic effects are expected to become negligible at large densities, as known, e.g., for colloidal glasses \cite{weeks2012colloidal-glass-transition} and confluent cell tissues \cite{manning2021jamming-and-arrest-of-cell-motion-in-biological-tissues, manning2016motility-driven-glass-and-jamming-transitions-in-biological-tissues}, and are absent in dry active systems \cite{chate2010vibrated-polar-disks, poeschel2010robots-move-collectively, dauchot2018spontaneously-flowing-crystal}.
The technical frugality allows us to cover the dilute and dense regimes in our numerical simulations, and to explore both  linear and  nonlinear response, across eight orders of magnitude in the shear rate.
The advantages of choosing harmonic spheres are threefold:
\begin{enumerate*}[label=\roman*)]
\item they have widely been employed as models for foams \cite{durian1995foam-mechanics-at-the-bubble-scale}, and, if endowed with activity, provide a useful model for biological tissues \cite{matoz2017nonlinear-rheology-in-a-model-biological-tissue, thirumalai2018cell-growth-rate-dictates-the-onset-of-glass};
\item the rheology of passive harmonic particles has been studied in detail \cite{ikeda2012unified, ikeda2013disentangling-glass-and-jamming, trulsson2015athermal}, thus allowing for a smooth connection with previous results to discriminate the new features brought about by activity;
\item a computational speed-up compared to hard spheres.
\end{enumerate*}

In our simulations, $N$ soft repulsive ABPs are placed at positions $\{\bm r_i\}_{i=1}^N$ in a $V=L^3$ cubic box with periodic boundary conditions.
They self-propel along their orientations $\bm n_i$ (with $|\bm n_i|=1$), with a speed $v_0$. 
Their otherwise overdamped Brownian dynamics obeys
\begin{align}
  \begin{split}
    \dot{\bm r}_i &= \mu\sum_{j\neq i}\bm F_{ij} + v_0\bm n_i + \dot{\gamma}y_i\bm e_x + \sqrt{2D_t}\bm \xi_i\,,\\
    \dot{\bm n}_i &= \sqrt{2D_r}\bm n_i\times\bm \nu_i\,.\\
  \end{split}
  \label{eq:eom}
\end{align}
The interaction forces derive from a harmonic repulsive pair potential $V(r)~=~\epsilon~(1-r/a)^2~\Theta(a-r)$, where $\Theta(r)$ denotes the Heaviside step function. 
To suppress crystallization we consider 50:50 bidisperse mixtures of $N=10^3$ particles with diameters $a$ and $\sqrt{2}\,a$, respectively \cite{ikeda2012unified, olsson2007critiical-scaling-of-shear-viscosity-at-the-jamming-transition}.
Both $\bm \xi_i$ and $\bm \nu_i$ are Gaussian white noises of zero mean and unit variance,  $D_t=\mu k_BT$ is the (bare) translational diffusion coefficient, and $D_r$ the rotational diffusivity fixed to $D_r=3D_t/a^2$.
With this choice of parametrization, the limit $v_0\to 0$ corresponds to a passive Brownian suspension, for which the fluctuation-dissipation theorem is satisfied.
We impose a linear velocity profile 
on the particles and apply Lees-Edwards boundary conditions 
(see Fig.~\ref{fig:phasediagram}(b))  \cite{lees1972computer, allen-tildesley2017computer-simulation-of-liquids}. 
The translational part of Eq.~\eqref{eq:eom} is integrated by an Euler-Mayurama scheme and the rotational part using the algorithm described in \cite{raible2004langevin-equation-rotation} 
(see \cite{SM} for details). 

Lengths are measured in units of the small particle diameter $a$, time $t$ in units of  $a^2/(\mu\epsilon)$ and temperature $T$ in units of $\epsilon/k_B$.
In the following, all observables are given in these units.
From Eq.~\eqref{eq:eom}, one can identify a set of nondimensional control parameters: 
the volume fraction $\phi$, the P{\'e}clet number $\mathrm{Pe}=v_0/a D_r$, quantifying activity, and the dimensionless shear rate $\dot{\gamma}$. 
We study the system at $T=10^{-6}\ldots 10^{-3}$ (most of the results presented are for $T=10^{-4}$), $\mathrm{Pe}=0,1, 3,10,30$, and a wide range of shear rates $\dot{\gamma}=10^{-7}\ldots 10$, including both the linear and nonlinear response regimes.
Note that the temperature unit allows to interpolate between soft and hard spheres, the latter being realized in the $T\to 0$ limit.
With this choice of parameters and constant $D_r$, the system remains homogeneous and avoids Motility-Induced Phase Separation, which is known to occur for monodisperse hard ABPs above a critical $\mathrm{Pe}\approx30$ \cite{wysocki2014cooperative-motion, turci2021phase-separation-and-multibody-effects-in-3D-ABP, omar2021phase-diagram-of-active-brownian-spheres}.
Also, the repulsive force is always several orders of magnitude larger than the self-propulsion to avoid a reentrant gas phase, as seen in \cite{fily2014freezing-and-phase-separation}.

In equilibrium ($\mathrm{Pe}=0$), as $\phi$ is increased, the system exhibits a dramatic slowing down of the dynamics that one identifies with a glass transition at $\phi_G$, characterized by the divergence of the viscosity and the emergence of a yield stress for $\phi>\phi_G$. As we show below, similar behavior is observed in the presence of activity $\mathrm{Pe}>0$, although in this case,  the location of the glass transition is shifted to higher densities $\phi_G(\mathrm{Pe})$, as previously reported for diverse models  \cite{berthier2013non, ni2013pushing-the-glass-transition, marchetti2011active-jamming, berthier2014nonequilibrium-glassy, levis2015single-particle-to-collective-effective-temperatures, berthier2017how-active-forces-influence-noneq-glass-transitions, paoluzzi2022mips-to-glassiness-in-dense-active-matter}. 

\begin{figure}[t]
  \centering
  \includegraphics[width=\linewidth]{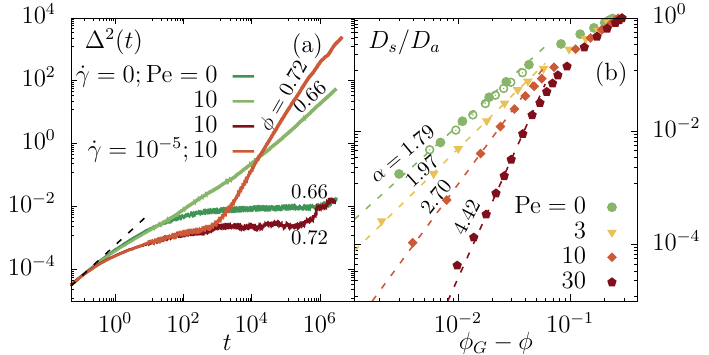}
  \caption{\label{fig:diff-coeff}
    Dynamic signatures of melting:
    (a) mean-square displacement $\Delta^2(t)$ at fixed $T=10^{-4}$,  $\mathrm{Pe}=0, 10$, for $\phi=0.66,\,0.72$, with and without shear, showing the melting of the  glass by activity (green curves) and by shear (red curves).
    The dashed line indicates the initial diffusive regime, $\Delta^2(t)=6D_tt$.
   (b) Long-time diffusion coefficients $D_s$ normalized by their ideal gas value $D_a$ at $T=10^{-4}$ (open green circles correspond to $T=10^{-3}$ and $\mathrm{Pe}=0$),  as a function of the distance to the critical density, $\phi_G - \phi$.
    Dashed lines are power law fits $D_s\propto (\phi_G - \phi)^\alpha$. 
  }
\end{figure}
We start our analysis by investigating the dynamics in the absence of shear by means of the Mean-Square Displacement (MSD), defined as $\Delta^2(t)=N^{-1}\sum_{i=1}^N\langle (\bm r_i(t) - \bm r_i(0))^2\rangle$.
The average is taken over different noise realizations. 
As shown in Fig.~\ref{fig:diff-coeff}(a), at $\phi=0.66$ the passive system exhibits caged dynamics, evidenced by the sub-diffusive (plateau) regime. For $\mathrm{Pe}=10$, particles diffuse in the same time window, showing that activity is able to fluidize the glass. 
We extract the long-time diffusion coefficient $D_s \equiv \lim_{t\to\infty}\Delta^2(t) / 6t$ in the range of parameters for which a diffusive regime, $\Delta^2(t)\propto t$, is observed. At high densities, $D_s(\phi)$ can be fitted by a power law  $D_s\propto (\phi_G - \phi)^\alpha$ that we use to locate the glass transition density $\phi_G(\mathrm{Pe})$ reported in Fig.~\ref{fig:phasediagram}(a) \cite{berthier2014nonequilibrium-glassy}.
Figure~\ref{fig:diff-coeff}(b) shows $D_s(\phi, \mathrm{Pe})$, normalized by the (active) ideal gas diffusion coefficient $D_a = D_t + \mathrm{Pe}^2D_r/6$ \cite{winkler2015virial}, as a function of $(\phi_G - \phi)$ for various $\mathrm{Pe}$. 
As activity primarily enhances diffusion, it moves the glass transition  from $\phi_G(0)=0.62$ at $\mathrm{Pe}=0$ to ever higher densites, up to $\phi_G(30)=0.72$ at $\mathrm{Pe}=30$ (see Fig.~\ref{fig:phasediagram}(a)). It might be tempting to simply interpret such a shift as resulting from an increase of the single particle effective temperature, defined by $T_{\text{eff}}=\mu D_a$, since the equilibrium glass transition density would also be shifted upon increasing $T$ \cite{berthier2009glass}. 
However, not only $\phi_G$ is affected by activity, but also the exponent $\alpha$.
It increases from $1.79$ at $\mathrm{Pe}=0$, 
up to $4.42$ at $\mathrm{Pe}=30$.
Since, in equilibrium, a slightly lower value ($\alpha=1.67$) is measured for the increased temperature  $T=10^{-3}$, the significant increase of $\alpha$ with $\mathrm{Pe}$ shows that activity cannot simply be reduced to an effective temperature, in this regime. 

Applying shear provides another route to fluidize the disordered solid state.
As shown in Fig.~\ref{fig:diff-coeff}(a), particles exhibiting caged dynamics in an active system at $\mathrm{Pe}=10$, $\phi=0.72>\phi_G(10)$, become mobile upon shearing.
The MSD displays super-diffusive, then diffusive behavior at long times.
We note that the obtained long-time diffusion is sensitive to finite-size effects: Lees-Edwards boundary conditions introduce a discontinuity in the shear profile, which becomes apparent in the MSD after a sufficiently long time (see \cite{SM} for details).

\begin{figure}[h]
  \centering
  \includegraphics[width=\linewidth]{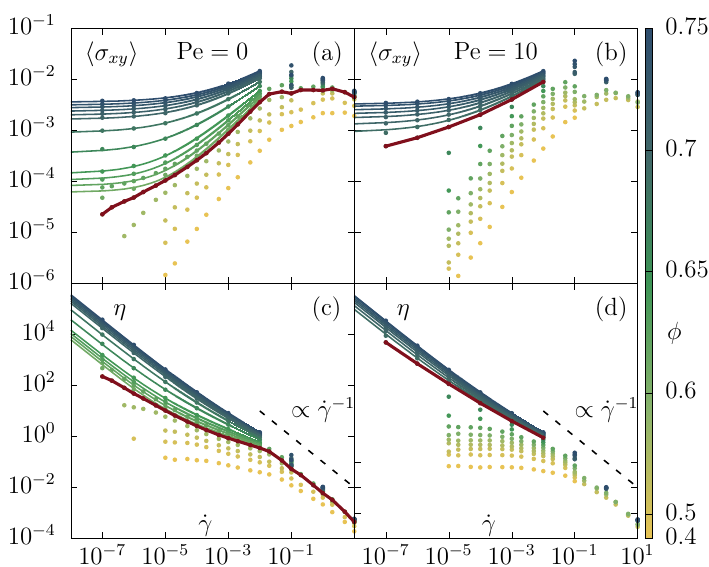}
  \caption{\label{fig:flow-curves}
    Flow curves for the shear stress $\langle\sigma_{xy}\rangle(\dot{\gamma})$ (top) and the viscosity $\eta(\dot{\gamma})$ (bottom).
    Color encodes $\phi$, the thick red lines correspond to $\phi_G(\mathrm{Pe})$ (as estimated via the diffusivity), and thin lines represent fits of the Herschel-Bulkley form $\sigma_{xy}(\dot{\gamma}) = \sigma_Y + (k\dot{\gamma})^n$, used to extract the yield stress $\sigma_Y$.
  }
\end{figure}

To characterize the flow properties, 
we measure the $xy$-component
$\sigma_{xy}(t) = -(2V)^{-1}\sum_{j\neq k}x_{ij}(t)F^y_{ij}(t)\,,$
of the stress tensor, using the Irving-Kirkwood expression \cite{irving1950statistical-mechanical-hydrodynamics}, 
from which we get the shear viscosity $\eta = \langle\sigma_{xy}\rangle/\dot{\gamma}$.
Activity contributes to the stress tensor with a self-term $\sigma^s_{\alpha\beta}=-V^{-1}\sum_ir^\alpha_iv_0n^\beta_i(t)$.  
As the orientations ${\bm n}_i$ are decoupled from the shear flow, $\sigma^s_{xy}$ fluctuates around zero  \cite{mandal2021shear-induced-orientational-ordering}.
The flow curves characterizing the rheology of the system at $T=10^{-4}$ and $\mathrm{Pe}=0,10$ are depicted in Fig.~\ref{fig:flow-curves}   (see \cite{SM} for more parameter values). 
For comparison with the rheology in the presence of activity, we reproduce the flow curves reported for the same system in the passive case, $\mathrm{Pe}=0$, before \cite{ikeda2012unified}.
In dense 2-dimensional active assemblies, shearing was observed to lead to orientational order at large persistence time \cite{mandal2021shear-induced-orientational-ordering}.
However, we did not find orientational correlations in our 3-dimensional model for the parameter range explored \cite{SM}.

At densities below $\phi_G(\mathrm{Pe})$, we find $\sigma_{xy}\propto \dot{\gamma}$ for small enough applied shear.
This corresponds to the Newtonian fluid regime, defining a linear viscosity $\eta_0 \equiv \lim_{\dot{\gamma}\to 0}\langle\sigma_{xy}\rangle/\dot{\gamma}$. 
In the nonlinear regime, at $\dot{\gamma} \gtrsim 10^{-2}$, we find shear-thinning in all cases, meaning that the shear flow reduces the  system's viscosity.
In this regime, we find a decay of the viscosity compatible with $\eta\sim 1/|\dot{\gamma}|$, as expected from Mode-Coupling Theory for  Brownian suspensions \cite{fuchs2002non-newtonian-viscosity-of-interacting-brownian-particles}.
At $\phi>\phi_G$, a finite yield stress $\sigma_Y\equiv\lim_{\dot{\gamma}\to 0} \sigma_{xy}(\dot{\gamma})$ appears, identified by a plateau in the stress flow curves. This results in a divergent viscosity, signalling the emergence of solidity. 
Since activity melts the solid, the system yields at higher densities at $\mathrm{Pe}=10$ compared to $\mathrm{Pe}=0$.

\begin{figure}[t]
  \centering
  \includegraphics[width=\linewidth]{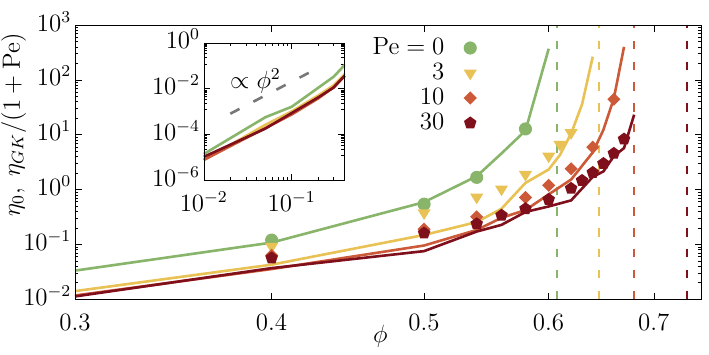}
  \caption{\label{fig:GK-viscosity}
    Density-dependent linear shear viscosity $\eta_0$, extracted from the flow curves (symbols), and $\eta_{GK}/(\mathrm{Pe}+1)$, from Green--Kubo (lines).
    Dashed vertical lines indicate $\phi_G(\mathrm{Pe})$.
    Inset: $\eta_{GK}\propto\phi^2$ at low densities .
  }
\end{figure}
  
In the fluid regime, the linear viscosity at low shear and $\mathrm{Pe}=0$ corresponds to the one given by the Green--Kubo (GK) relation
$
  \eta_{GK} = \frac{V}{k_BT}\int_0^\infty\mathrm dt\,\langle\sigma_{\alpha\beta}(t)\sigma_{\alpha\beta}(0)\rangle_0\,,
$
for $\alpha\neq\beta$, where $\langle*\rangle_0$ denotes an average over the unperturbed ($\dot{\gamma}=0$) equilibrium distribution. As shown in  Fig.~\ref{fig:GK-viscosity}, the shear viscosity $\eta_0$ extracted from the low $\dot{\gamma}$ plateau in the flow curves in Fig.~\ref{fig:flow-curves}(c) matches $\eta_{GK}$,  measured from the GK relation by direct integration of the equilibrium stress correlation function. In the presence of activity,  GK relations do not need to hold anymore, although  extensions of linear response theory to active systems have recently been proposed  \cite{cengio2019linear-response, solon2019response, cengio2021fluctuation-dissipation, dadhichi2022time}, providing GK relations involving  steady-state correlation functions \cite{cengio2019linear-response, cengio2021fluctuation-dissipation}.
Here we apply the same procedure in the presence of activity, thus replacing  equilibrium by  steady-state stress correlations, and find a good agreement between $\eta_0$ and $\eta_{GK}$ if we replace $T$ in the GK expression by an effective temperature $T_{\mathrm{eff}}=T(\mathrm{Pe}+1)$, which is again different from the ideal-gas expectation $T_{\mathrm{eff}}=\mu D_a$, see Fig.~\ref{fig:GK-viscosity}.
In all cases, $\eta_0$ increases with $\phi$ and eventually diverges at $\phi_G$, providing yet another estimate for the onset of solidity. 
In dilute conditions, we find that the GK viscosity  grows like $\sim\phi^2$ (inset of Fig.~\ref{fig:GK-viscosity}), as predicted for dilute Brownian suspensions: $\eta/\eta_0 = 1 + 2.5\phi + 7.6\phi^2$  \cite{batchelor1972determination-of-bulk-stress-in-a-suspension-of-spherical-particles}. 
We only observe the $\phi^2$ contribution, as there is no solvent in our system. 

\begin{figure}[t]
  \centering
  \includegraphics[width=\linewidth]{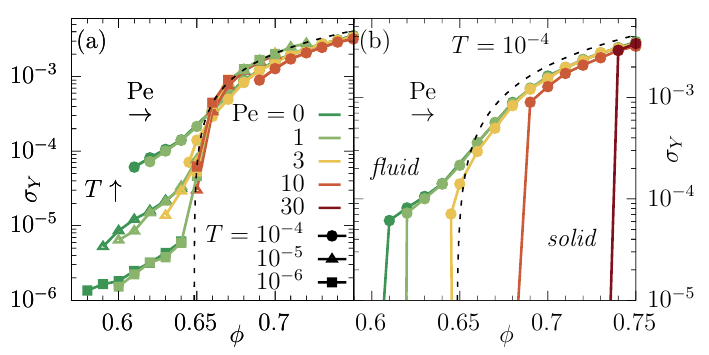}
    \caption{\label{fig:yield-stress}
      (a) Yield stress $\sigma_Y$ as a function of $\phi$ for different $\mathrm{Pe}$ (color) and $T$ (symbols):
      An increase in $T$ raises $\sigma_Y$ while, on the contrary, increasing $\mathrm{Pe}$  lowers it. 
      The dashed line  given by $(\phi - \phi_J)^{\alpha'}$ with $\phi_J= 0.648$ and $\alpha'= 1.04$, marks the athermal jamming limit.
      (b) Yield stress $\sigma_Y(\phi)$ at $T=10^{-4}$ for the parameters used to construct the yield surface in Fig.~\ref{fig:phasediagram}.
      Steep lines connect to  $\phi_G(\mathrm{Pe})$.
  }
\end{figure}

Above $\phi_G$ the viscosity diverges and the system acquires a yield stress $\sigma_Y$ that we measure by fitting the Herschel-Bulkley relation, $\sigma_{xy}(\dot{\gamma}) = \sigma_Y + (k\dot{\gamma})^n$ \cite{herschel-bulkley1926konsistenzmessungen-von-gummi-benzolloesungen}, to the flow curves, see Fig.~\ref{fig:flow-curves}. We report  the obtained  yield stress  $\sigma_Y(\phi)$ as a function of volume fraction for different values of $\mathrm{Pe}$ and $T$ in  Fig.~\ref{fig:yield-stress}.
At $\mathrm{Pe}=0$ and finite $T$,  the emergence of solidity is controlled by the glass transition at a critical density $\phi_G(T)$ that increases with $T$.
At $T=0$, it is instead controlled by the jamming transition at $\phi_J\approx 0.648$. 
Both glass and jamming physics affect the shape of the yield surface $\sigma_Y (\phi,T)$. 
At $T\leqslant 10^{-4}$, for which $\phi_G<\phi_J$, the system displays two distinct rheological regimes associated to glassy and jamming physics. 
First, $\sigma_Y$ increases gently with $\phi$ and $T$ up to $\phi\approx\phi_J$ (dashed lines in Fig.~\ref{fig:yield-stress}). Above this value, the behavior of $\sigma_Y$ changes qualitatively: it grows faster with $\phi$ close to $\phi_J$, with little $T$-dependence, following $\sigma_Y\sim(\phi-\phi_J)^{\alpha'}$ \cite{ikeda2012unified}, see  Fig.~\ref{fig:yield-stress}(a). 
%

 In the presence of activity, $\mathrm{Pe}>0$, and at finite $T$, $\sigma_Y$ displays a $T$-sensitive glass-like branch for $\phi<\phi_J$ followed by the $T$-insensitive jamming branch, see  Fig.~\ref{fig:yield-stress}(a). 
 The yield stress curves do not follow the trend one would expect if activity could be subsumed into an extra source of noise and encoded by an increased effective temperature, although both $\mathrm{Pe}$ and $T$ shift the fluid-solid transition to higher densities.
 As $T$ increases, $\sigma_Y$ increases, shifting the curves in Fig.~\ref{fig:yield-stress}(a) upwards.
In contrast, increasing $\mathrm{Pe}$ slightly lowers the yield stress, shifting the curves towards
      the right and quickly collapsing them into the jamming line.
For $\mathrm{Pe}>3$,  $\phi_G(\mathrm{Pe})>\phi_J$, and  a finite yield stress can only emerge in the parameter regime controlled by jamming \cite{olsson2007critiical-scaling-of-shear-viscosity-at-the-jamming-transition}, where $\sigma_Y\sim(\phi-\phi_J)^{\alpha'}$ universally applies, independently of $\mathrm{Pe}$. 
The crossover between glass and jamming rheology can thus be tuned by activity and is eventually lost, as  it pushes $\sigma_Y$ towards the athermal limit. 
The separation between $\phi_G$ and $\phi_J$ progressively vanishes as activity increases. 
An overview over the impact of activity on the yield surface at $T=10^{-4}$ is represented in the fluid-glass-jamming phase diagram in Fig.~\ref{fig:phasediagram}.

In summary, we have studied soft ABPs under shear, from the fluid to disordered solid regime. 
The Newtonian fluid viscosities in the zero-shear limit are compatible with those obtained from the Green-Kubo relation, once the Brownian temperature $T$ is replaced by an effective temperature $T_{\text{eff}}$ that depends on the active P\'eclet number.
This strategy follows earlier ideas to quantify the violations of the fluctuation-dissipation theorem in active systems. In the dilute limit,  $T_{\text{eff}}\sim \mathrm{Pe}^2$. At intermediate densities $T_{\text{eff}}$  generically depends on the observables used to define it \cite{levis2015single-particle-to-collective-effective-temperatures, cengio2019linear-response, petrelli2020effective}. 
As the packing fraction is increased towards the fluid-solid transition, the diffusivity decays according to a critical power-law $D_s\sim(\phi_G-\phi)^{\alpha}$, with  $\alpha$ increasing from $\alpha\approx 1.8$ for $\mathrm{Pe}=0$, to $\alpha\approx 4.4$ for $\mathrm{Pe}=30$, a behavior hardly interpretable on the grounds of an effective temperature anymore.
The  glass-jamming phase diagram (Fig.~\ref{fig:phasediagram}) reveals that ABP rheology in the solid regime is mainly controlled by jamming. Although both $T$ and $\mathrm{Pe}$ push $\phi_G$ to higher values, activity, as opposed to temperature, eases the yielding, and, in this respect, could rather play a role similar to shear, once the persistence length of the active motion exceeds the typical cage length \cite{morse2021link-active-matter-sheared-granular}.  
Our work thus calls for further theoretical efforts to better understand the fundamental role played by   non-equilibrium fluctuations introduced by activity in dense disordered systems, for instance, by extending elastoplastic models of yielding to account for self-propulsion or analyzing dynamic heterogeneities in active glasses \cite{yamamoto1998dynamics, voigtmann2014nonlinear, nicolas2018deformation, kamani2021unification}.
By providing the first quantitative jamming phase diagram of an active system
it should serve as a helpful reference for future studies of soft active assemblies.
The fact that their rheology constitutes a separate class from the known pressure-controlled flows of hard spheres \cite{boyer2011suspension-and-granular-rheology} brings us a step closer to reconciling the latter with
observations in dense assemblies of cells, for which only qualitative  phase diagrams have been sketched so far \cite{mongera2018fluid-to-solid-jamming-transition,manning2021jamming-and-arrest-of-cell-motion-in-biological-tissues, lenne2022sculpting-tissues-by-phase-transitions, campas2022nuclear-jamming-transition-in-vertebrate-organogenesis}.
It should also provide a good starting point for future attempts to elucidate the relationship between the glass and jamming transitions in biological systems.

\paragraph{Acknowledgments}
We warmly thank T. Voigtmann, I. Pagonabarraga, S. Dal Cengio, A. Ikeda and L. Berthier for useful exchanges. 
D.L. acknowledges MCIU/ AEI/FEDER for financial support under Grant Agreement No. RTI2018-099032-J-I00. 

\bibliographystyle{apsrev4-2}
\bibliography{refs}

\end{document}